\documentclass{iau}
\usepackage{graphicx}

\newcommand{\mytilde}{\raise.17ex\hbox{$\scriptstyle\mathtt{\sim}$}}

\title[The Value of $H_0$ from Gaussian Processes] 
{The Value of $H_0$ from Gaussian Processes}

\author[Vinicius C. Busti, Chris Clarkson \& Marina Seikel]   
{Vinicius C. Busti$^1$, Chris Clarkson$^1$
 \and Marina Seikel$^{2,1}$}

\affiliation{$^1$ Astrophysics, Cosmology \& Gravity Centre (ACGC), and Department of Mathematics and Applied Mathematics, University of Cape Town, Rondebosch 7701, Cape Town, South Africa  \\ email: {\tt vinicius.busti@iag.usp.br} \\[\affilskip]
$^2$ Physics Department, University of Western Cape, Cape Town 7535, South Africa}

\pubyear{2014}
\volume{306}  
\pagerange{119--126}
\jname{Statistical Challenges of 21$^{st}$ Century Cosmology}
\editors{A. Heavens, J.-L. Starck \& A. Krone-Martins, eds.}
\begin{document}

\maketitle

\begin{abstract}
A new non-parametric method based on Gaussian Processes (GP) was proposed recently to measure the Hubble constant $H_0$. The freedom in this approach comes in the chosen covariance function, which determines how smooth the process is and how nearby points are correlated. We perform coverage tests
with a thousand mock samples within the $\Lambda$CDM model in order to determine what covariance function provides the least biased results. The function Mat\'{e}rn(5/2) is the best with sligthly higher errors than other covariance functions, although much more stable when compared to standard parametric analyses.

\end{abstract}

\firstsection 
\section{Introduction}

The difference between the value of the Hubble constant $H_0$ determined by {\it Planck} \cite[(Planck Collaboration 2013)]{planck} and by local measurements \cite[(Riess et al. 2011)]{riess} shows a 2.3$\sigma$ tension.
In order to understand what could generate such a discrepancy, many attempts in the literature were done searching for new physics or systematic errors 
(e.g. \cite[Marra et al. 2013]{marra}, \cite[Spergel et al. 2013]{spergel}, \cite[Efstathiou 2014]{efs}, \cite[Wyman et al. 2014]{neutrinos}, 
\cite[Holanda et al. 2014]{holanda}, \cite[Clarkson et al. (2014)]{clarkson}).

Recently, we proposed a new method to determine $H_0$ by applying {\it Gaussian Processes} (GP), which is a non-parametric procedure, to reconstruct $H(z)$ data and extrapolating to redshift zero \cite[(Busti et al. 2014)]{bcs2014}. We selected 19 $H(z)$ measurements (\cite[Simon et al. 2005]{simon2005}, \cite[Stern et al. 2010]{stern2010}, \cite[Moresco et al. 2012]{moresco2012}) based on cosmic chronometers and obtained $H_0= 64.9 \pm 4.2$ km s$^{-1}$ Mpc$^{-1}$, which is compatible with {\it Planck} but shows a tension with local measurements.

Here, we use mock samples in order to test our method. Basically, we are interested to see which covariance function adopted in the GP analysis provides the best match with the fiducial cosmological model. As we shall see, the Mat\'{e}rn(5/2) performs better, with the standard squared exponential covariance function underestimating the errors. In the next section GP will be briefly described, showing our results in Sec. 3. 
We draw our conclusions and discuss future improvements in Sec. 4.

\section{Gaussian Processes (GP)}

A gaussian process allows one to reconstruct a function from data without assuming a parametrisation for it. While a gaussian distribution is
a distribution over random variables, a gaussian process is a distribution over functions. We use GaPP (Gaussian Processes in 
Python)\footnote{http://www.acgc.uct.ac.za/\mytilde seikel/GAPP/index.html} \cite[(Seikel et al. 2012))]{gapp} in order to reconstruct the Hubble parameter as a function of the redshift.

Basically, the reconstruction is given by a mean function with gaussian error bands, where the function values at different points $z$ and $\tilde{z}$ are connected through a 
covariance function $k(z,\tilde{z})$. This covariance function depends on a set of hyperparameters.  For example, the general purpose squared 
exponential (Sq. Exp.) covariance function is given by
\begin{equation}
 k(z,\tilde{z}) = \sigma_f^2 \exp\left\{-\frac{(z-\tilde{z})^2}{2l^2}\right\}.
\end{equation}
In the above equation we have two hyperparameters, the first $\sigma_f$ is related to typical changes in the function value while the second $l$ is related to the distance one needs
to move in input space before the function value changes significantly.
We follow the steps of \cite{gapp} and determine the maximum likelihood value for $\sigma_f$ and $l$ in order to obtain the value of the function. In this way, we are able to 
reconstruct the Hubble parameter as a function of the redshift from $H(z)$ measurements.  Many choices of covariance function are possible, and we consider a variety below.

\section{Results}

The freedom in the GP approach comes in the covariance function. While in traditional parametric analyses we choose a model to characterise what is our prior belief
about the function in which we are interested, with GP we ascribe in the covariance function our priors about the expected function properties (e.g. smoothness, correlation scales etc.).

We consider the Sq. Exp. covariance function and three examples from the Mat\'{e}rn family:

\begin{equation}
k(z,\tilde{z}) = \sigma_f^2 \frac{2^{1-\nu}}{\Gamma(\nu)} \left( \frac{\sqrt{2\nu(z-\tilde{z})^2}}{l} \right)^{\nu} K_{\nu}\left( \frac{\sqrt{2\nu(z-\tilde{z})^2}}{l} \right),
\end{equation}
where $K_{\nu}$ is a modified Bessel function and we choose $\nu=5/2$, $7/2$ and $9/2$ (see \cite[Seikel \& Clarkson 2013 for more discussions]{sc2013}).
Writing $\nu = p + 1/2$, each Mat\'{e}rn function is $p$ 
times differentiable as are functions drawn from it, and
the squared exponential is recovered for $\nu \rightarrow \infty$. Increasing $\nu$ increases the width of the covariance function near the peak implying stronger correlations from 
nearby points for a fixed correlation length $\ell$. For comparison purposes, we also consider two standard parametric models: a flat $\Lambda$CDM 
model and a flat XCDM model.

The results are shown in Table \ref{table1} together with the constraints from the 19 $H(z)$ data. The coverage test of each covariance function and parametric model was performed by creating 1000 mock catalogues of 19 data points with the 
same redshifts and error-bars of the measured points
in a fiducial $\Lambda$CDM model. For each model realisation a value of $H_0$ was derived.  The third and fourth columns
of Table \ref{table1} show the frequency the true value for $H_0$ was recovered inside the $1\sigma$ and $2\sigma$ regions. So, for example, the Mat\'{e}rn$(9/2)$ covariance function 
captures the true value at 1(2)$\sigma$ about 60\%(94\%) of the time~-- alternatively, the 1(2)$\sigma$ region should be interpreted as a 60\%(94\%) confidence interval. This provides 
a way to re-normalise the $n\sigma$ intervals for a given covariance function and a prior model assumption, which we show between parentheses for 1$\sigma$(68\%) errors in 
Table \ref{table1}. 
Therefore, this is an attempt to quantify a possible systematic error from the covariance functions \emph{assuming the true model 
is $\Lambda$CDM}. We also considered some different fiducial models with a time-varying dark energy equation of state, 64 data points in the redshift range $0.1 < z < 1.8$, with coverages
showing the same pattern as depicted in Table \ref{table1}.
It is important to note this is a model-dependent comparison which relies on the knowledge of the true model in advance, which is never the case, and 
changes with the quality of the data. The coverage can change with a different underlying model as well~-- but note that the errors are actually much more stable than switching 
from $\Lambda$CDM to XCDM.

\begin{table}
\caption{$H_0$ constraints from 19 $H(z)$ measurements.}
\label{table1}
\begin{center}
\begin{tabular}{@{}ccccc@{}}
\hline Method & $H_0$ $\pm$  $1\sigma$ &  Coverage & Coverage   \\
          & (km s$^{-1}$ Mpc$^{-1}$)   & 1$\sigma$ & 2$\sigma$     
\\ \hline\hline
\\

Sq. Exp.                    &  64.9 $\pm$    4.2(5.9)  &  0.527  &  0.905     \\
Mat\'{e}rn$(9/2)$           &  65.9 $\pm$    4.5(5.6)  & 0.594   &  0.939     \\
Mat\'{e}rn$(7/2)$           &  66.4 $\pm$    4.7(5.7)  &  0.610  &  0.946    \\
Mat\'{e}rn$(5/2)$           &  67.4 $\pm$    5.2(5.5)  & 0.665   &  0.959    \\
$\Lambda$CDM                &  68.9 $\pm$    2.8  & 0.676   &  0.938   \\
XCDM                        &  69.0 $\pm$    6.7  & 0.685   &  0.939   \\

\hline
\end{tabular}
\end{center}
\end{table}

\section{Conclusions}

We have applied GP to reconstruct $H(z)$ data and from it extrapolate to redshift zero to obtain $H_0$ \cite[(Busti et al. 2014)]{bcs2014}. Based on a set of 1000 mock samples, we have tested the method assuming a fiducial flat $\Lambda$CDM model by considering four different covariance functions and applying a coverage test. We have shown Mat\'{e}rn(5/2) represents better the errors, with errors slightly higher than the other covariance functions. A heuristic method to recalibrate the errors for different covariance functions was also provided within the $\Lambda$CDM model.

Possible improvements can be achieved by marginalizing over the hyperparameters and comparing the results using Bayesian model comparison tools, which will allow a direct assessment of performance with no need to rely on a fiducial model.


\begin{thebibliography}{}

\bibitem[Busti et al. (2014)]{bcs2014}
{Busti, V. C., Clarkson, C. \& Seikel, M.} 2014,
\textit{MNRAS} (Letters) 441, L11 [preprint(arXiv:1402.5429)]


\bibitem[Clarkson et al. (2014)]{clarkson}
{Clarkson, C., Umeh, O., Maartens, R. \& Durrer, R.} 2014,
preprint(arXiv:1405.7860) 

\bibitem[Efstathiou (2014)]{efs}
{ Efstathiou, G.} 2014,
\textit{MNRAS}, 440, 1138 

\bibitem[Holanda et al. (2014)]{holanda}
{Holanda, R. F. L., Busti, V. C. \& Pordeus da Silva, G.} 2014,
\textit{MNRAS} (Letters), 443, L74 [preprint(arXiv:1303.5076)] 

\bibitem[Marra et al. (2013)]{marra}
{Marra, V., Amendola, L., Sawicky, I., \& Walkenburg, W.} 2013,
\textit{Phys. Rev. Lett.}, 110, 241305 

\bibitem[Moresco et al. (2012)]{moresco2012}
{Moresco, M. et al.} 2012,
\textit{J. Cosmol. Astropart. Phys.}, 8, 6


\bibitem[Planck Collaboration (2013)]{planck}
{Planck Collaboration, Ade, P. A. R. et al.} 2013,
preprint(arXiv:1303.5076) 

\bibitem[Riess et al. (2011)]{riess}
{Riess, A. G. et al.} 2011, 
\textit{ApJ}, 730, 119

\bibitem[Seikel et al. (2012)]{gapp}
{Seikel, M., Clarkson, C. \& Smith, M.} 2012,
\textit{J. Cosmol. Astropart. Phys.}, 6, 36


\bibitem[Seikel \& Clarkson (2013)]{sc2013}
{Seikel, M. \& Clarkson, C.} 2013,
preprint(arXiv:1311.6678)


\bibitem[Simon et al. (2005)]{simon2005}
{Simon, J., Verde, L. \& Jimenez, R.} 2005,
\textit{Phys. Rev. D}, 71, 123001

\bibitem[Spergel et al. (2013)]{spergel}
{Spergel, D., Flauger, R. \& Hlozek, R.} 2013,
preprint(arXiv:1312.3313) 

\bibitem[Stern et al. (2010)]{stern2010}
{Stern, D., Jimenez, R., Verde, L., Kamionkowski, M. \& Stanford, S. A.} 2010,
\textit{J. Cosmol. Astropart. Phys.}, 2, 8


\bibitem[Wyman et al. (2014)]{neutrinos}
{Wyman, M., Rudd, D. H., Vanderveld, A. \& Hu, W.} 2014,
\textit{Phys. Rev. Lett.}, 112, 051302 

\end{thebibliography}
\end{document}